\documentclass[12pt]{iopart}
\usepackage{iopams}
\usepackage{graphicx}

\newcommand{\tra}{\mathrm{Tr}}
\newcommand{\be}{\begin{equation}}
\newcommand{\ee}{\end{equation}}
\newcommand{\ben}{\begin{eqnarray}}
\newcommand{\een}{\end{eqnarray}}

\begin{document}
\title[Entanglement quantification from incomplete measurements]{Entanglement quantification from incomplete measurements:
Applications using photon-number-resolving weak homodyne detectors}

 \author{Graciana Puentes}
 \address{Clarendon Laboratory, Department of Physics, University of Oxford, OX1 3PU Oxford, UK\\
 Email : g.puentes1@physics.ox.ac.uk}

\author{Animesh Datta}
 \address{Institute for Mathematical Sciences, Imperial College London,
 London SW7 2PG, UK}
 \address{QOLS, The Blackett Laboratory, Imperial College London,
 London SW7 2BW, UK}

\author{Alvaro Feito}
 \address{Institute for Mathematical Sciences, Imperial College London,
 London SW7 2PG, UK}
 \address{QOLS, The Blackett Laboratory, Imperial College London,
 London SW7 2BW, UK}

\author{Jens Eisert}
  \address{Institut f{\"u}r Physik, Universit{\"a}t Potsdam,
  D-14469 Potsdam, Germany}
    \address{Institute for Advanced Study Berlin,
    D-14193 Berlin, Germany}
 \address{Institute for Mathematical Sciences, Imperial College London,
 London SW7 2PG, UK}

\author{Martin B.\ Plenio}
 \address{Institut f{\"u}r Theoretische Physik, Universit{\"a}t Ulm, D-89069 Ulm, Germany}
 \address{Institute for Mathematical Sciences, Imperial College London, London SW7 2PG, UK}
 \address{QOLS, The Blackett Laboratory, Imperial College London, London SW7 2BW, UK}

\author{Ian A.\ Walmsley}
 \address{Clarendon Laboratory, Department of Physics, University of Oxford, OX1 3PU Oxford, UK}

\begin{abstract}
The certificate of success for a number of important quantum information processing protocols, such as  entanglement distillation, is based on the difference in the entanglement content of
the quantum states before and after the protocol. In such cases, effective
bounds need to be placed on the entanglement of non-local states
consistent with statistics obtained from local measurements.
In this work, we study numerically the ability of a novel type of homodyne detector which combines phase sensitivity and photon-number resolution to set accurate bounds on
the entanglement content of two-mode quadrature squeezed states
without the need for full state tomography. We show that it is
possible to set tight lower bounds on the entanglement of
a family of two-mode degaussified states using only a few measurements. This presents a significant improvement over  the
resource requirements for the experimental demonstration of
continuous-variable entanglement distillation, which traditionally
relies on full quantum state tomography.
\end{abstract}

\maketitle

\section{Introduction}
\label{sec:Intro}

Entanglement is a fundamental characteristic of quantum systems and a primary resource in quantum information science. Therefore methods to experimentally measure the entanglement of the quantum state of a system are important both for the interpretation of experiments involving quantum systems and for verifying the operation and capacity of a quantum processor or communications system. The most common approach to this problem is to perform quantum tomography of the unknown state of the system~\cite{dariano03}. Quantum state tomography amounts to measuring a tomographically
complete set of observables, followed by suitably postprocessing
the data.  For example, in systems specified by continuous variables (such as the quadrature amplitudes of an optical field, or the position and momentum of a mechanical oscillator), the basic
theoretical principle is that a collection of probability
distributions of the transformed continuous variables is the
Radon transform of its Wigner function.  Starting from
experimentally measured marginals, therefore, an inverse Radon
transform gives the Wigner function from which elements of the
density matrix can be obtained. The notion was first
experimentally realized in the domain of quantum
optics~\cite{Raymer, Dunn}. Since then quantum state tomography has been
improved to give controlled statistical errors using
maximum-likelihood or least squares \cite{LSE}, made more
efficient for low-rank states using ideas of compressed sensing
\cite{CS}, and equipped with statistical error bars~\cite{as09}.
This is of particular importance in the case of density matrices
of non-classical states, which are typically characterized by a
negative quasi-probability distribution, such as the Wigner
function~\cite{Wigner}. Reconstruction of such non-classical
states is indeed a part and parcel of experimental demonstrations
of quantum information protocols. Non-classical features may be difficult to reconstruct. In photonic applications, this is often due to low quantum detection efficiencies, leading to noisy measurements.~\cite{lr09,Ourjoumtsev07}  Typically, overall detection efficiencies above $50 \%$ are required.  However, direct detection of other non-classical signatures may be effected using different sorts of detectors. For example, weak-field homodyne detection coupled with photon counting provides a means to detect entanglement in Gaussian states.~\cite{Grangier88, Kuzmich00}

In this work we present an extensive numerical study of a strategy that 
provides robust direct quantative estimates for the entanglement
content of a state, without the need of full quantum state
tomography. In order to accomplish this task, we systematically
investigate the performance of a weak-field homodyne detector with
photon-number resolution as an experimentally feasible component
for the construction of local measurement operators. These will be
a set of positive operator valued measurements (POVMs). The POVM
elements required for such a construction are characterized by a
model of the homodyne detector, based on a previous
characterization of the time-multiplexed photon-number-resolving
detectors~\cite{Puentes}. This detector has also been
characterized using the nascent field of detector
tomography~\cite{detectomo}. The fundamental question we will
answer here is the entanglement content of the least entangled
state consistent with the available measurement
data~\cite{ap06,eba07,grw07,p09}. Thus, we will be left with a
\textit{lower} bound on the entanglement of the state in question.
In particular, this procedure can be used for setting a lower
bound on the Logarithmic Negativity \cite{p05}, the evaluation of
which can be reduced to an efficiently solvable class of convex
optimization problems called semidefinite
programs~\cite{ap06,eba07,grw07,p09,bv05}. We apply our technique
to two mode photon-subtracted quadrature squeezed states. Setting
bounds on such a family of non-Gaussian quantum states is of major
significance for the implementation of a continuous-variable
entanglement distillation protocol~\cite{Eisert,distillation}.

Although we will primarily be concerned with continuous-variable
entanglement distillation~\cite{distillation} in this work, we
must make it clear at the outset that the technique studied here
can be applied to any task that aims to manipulate entanglement
between spatially separated observers by local operations and
classical communications (LOCC), and subsequently confirm the
outcome, also by means of local operations and classical
communications. This goes back to the resource nature of
entanglement, and the ability to manipulate it by LOCCs.
Continuous variable entanglement distillation is an important
instance of such a situation. It should also be clear that we are not
limited to entirely optical settings, and similar techniques should
be helpful to eventually identify entanglement in opto-mechanical
settings, say of entanglement between a micromirror and an
optical mode.

In cases such as these, it is often possible to gather enough
information by a limited number of measurements to assess the
correlations in the state. The natural question then is whether
the correlations revealed by these local measurements (aided
possibly by classical communication) represent classical
correlations, or entanglement~\cite{ap06,p09}. This circumvents
the necessity of the resource-intensive process of quantum state
tomography. The method is also more robust with respect to measurement
errors than full state tomography. Importantly,
no a-priori assumptions concerning the purity or the
specific form of the states enter the certifiable bound on the
degree of entanglement.

The paper is structured as follows. In Sec.~(\ref{sec:SDPs}), we
formalize as a semidefinite program (SDP) the problem of putting
lowers bounds on the entanglement content of states using
localized measurement statistics. Sec.~(\ref{sec:PNR}) describes
the specific time-multiplexed homodyne detector that we use to
build these localized measurements. In Sec.~(\ref{sec:CVentdist}),
we present the numerical results on the bounds set on the
entanglement content of a two-mode photon subtracted quadrature
squeezed state, for different values of relevant experimental
parameters. We also present an extensive numerical exploration of
the performance of the detector under different experimental
conditions. In particular, we analyze the required phase accuracy
and phase stability in our homodyne scheme. We also discuss the
tolerance of the convex optimization algorithm to experimental
noise. Finally, in Sec.~(\ref{sec:conc}) we report the
conclusions. As a matter of notation, all logarithms in this paper
are taken to base $2$.

\section{Lower bounds using convex optimization}
\label{sec:SDPs}

As stated in the introduction, we are seeking the amount of
entanglement in the least entangled state compatible with a set of
measurement results. Mathematically, this can be presented as
 \be
 \label{eq:Emin}
 E_{\min} = \min_{\rho}\{E(\rho): \tra(\rho
 M_i)=m_i\},
 \ee
where $E$ is the measure of entanglement, and $M_i$ are the
measurements made with measurement data $m_i.$ Additional
constraints that $\rho$ is a density matrix, i.e., positive, and
$\tra(\rho)=1$, are also imposed. The latter is easily done by
setting $M_0=\mathbb{I}$ and $m_0=1.$ Depending on the measure of
entanglement, and the measurements chosen, the minimization in Eq.\
(\ref{eq:Emin}) can even be performed analytically, but generically
that is not the case. Here, we briefly present a technique following
the presentation in Refs.\ \cite{ap06,eba07} that allows
the above problem to be cast as a semidefinite program when the
entanglement measure is the Logarithmic Negativity~\cite{p05}.

Logarithmic Negativity is defined as the logarithm of the 1-norm
of the partial transposed density matrix $\|\rho^{T_1}\|_1.$ The
1-norm can be expressed as~\cite{b97}
 \be
\|\rho^{T_1}\|_1=\max_{\|H\|_{\infty}=1}\tra(H\rho^{T_1}) =
\max_{\|H\|_{\infty}=1}\tra(H^{T_1}\rho),
 \ee
with the maximization being over all hermitian operators $H$, where
$\|.\|_\infty$ denotes the standard matrix operator norm, namely the
largest singular value of the matrix.
Using the monotonicity of the logarithm, the minimization
in Eq.\ (\ref{eq:Emin}) can be rewritten as
 \be
 \label{eq:minmax}
 \mathcal{N}_{\min} = \log\min_{\rho}\left\{ %%@
\max_{H}\{\tra(H^{T_1}\rho)\big| \|H\|_{\infty}=1\} : \tra(\rho
 M_i)=m_i\right\}.
 \ee
The minimax equality allows us to interchange the maximization and
the minimization, leading to
 \be
 \label{eq:maxmin}
 \mathcal{N}_{\min} = \log\max_{H}\left\{\min_{\rho} %%@
\{\tra(H^{T_1}\rho): \tra(\rho
 M_i)=m_i\}: \|H\|_{\infty}=1 \right\}.
 \ee
 For any real numbers $\{\nu_i\}$ for which
 \begin{equation}
        H^{T_1}\geq\sum_i\nu_iM_i,
 \end{equation}
 clearly the lower bound
 \begin{equation}
        \tra(H^{T_1}\rho)\geq \sum_i\nu_i \tra(M_i\rho) = \sum_i\nu_i m_i.
 \end{equation}
holds true for states $\rho$.
%Using Lagrange multipliers $\nu_i,$ the inner minimization amounts
%to minimizing $\tra[\rho(H^{T_1}-\sum_i\nu_iM_i)] + \sum_i
%\nu_im_i,$ where $\rho$ is a positive operators. If
%$(H^{T_1}-\sum_i\nu_iM_i)$ has negative eigenvalues, the minimum
%is $-\infty$, making the outer maximization facile. We thus
%restrict $(H^{T_1}-\sum_i\nu_iM_i) \geq 0,$ in which case the
%minimum $\sum_i \nu_im_i$ is obtained for $\rho=0.$
Thus we get
 \be
 \label{eq:maxmax}
 \mathcal{N}_{\min} \geq \log\max_{H}\left\{\max_{\nu_i} \big\{\sum_i %%@
\nu_im_i: H^{T_1}\geq\sum_i\nu_iM_i \big\} : \|H\|_{\infty}=1 %%@
\right\}.
 \ee
Note that the state $\rho$ drops completely out of contention now.
Since the inner minimization in Eq.\ (\ref{eq:maxmin}) is a
semidefinite program, strong duality in the strictly feasible case
ensures equality in Eq.\ (\ref{eq:maxmax}). Thus, having fixed the
operators that we choose to measure $M_i,$ any choice of $H$ and
$\nu_i$ such that $H^{T_1}\geq\sum_i\nu_iM_i$ and
$\|H\|_{\infty}=1,$ provides us with a lower bound on the
Logarithmic Negativity of states which provide expectation values
of $m_i.$ Finally, we can rewrite Eq.\ (\ref{eq:maxmax}) as
 \ben
 \label{eq:sdp}
 &&\mathrm{maximize}\;\;\;\log\Big(\sum_i \nu_im_i\Big), \\
 &&\mathrm{subject\;to}\;\;H^{T_1}\geq\sum_i\nu_iM_i,\nonumber\\
 && \;\;\;\;\;\; \mathrm{and}\;\;\;   -\mathbb{I} \leq H \leq %%@
\mathbb{I},\nonumber
 \een
which can be solved quite easily using standard SDP solvers, like
SeDuMi~\cite{sedumi}, once we have decided what our measurements
$M_i$ are. Since these are to be local, the typical form of the
measurement, in the case of bipartite states, is
 \be
 \label{eq:localmeas}
M_i = \Pi^{1}_j\otimes\Pi^{2}_k.
 \ee
 The problem is thus reduced to the construction of the operators
$\Pi^{1,2}_j,$ which is what we move onto in the next section. In
passing, we mention that the choice of these measurement operators
can also be cast as a SDP, although it is more challenging to
incorporate the locality constraint into its framework.

This idea gives useful and practically tight bounds to the
entanglement content, not having to assume any a-priori knowledge
about the state, or properties of it such as its purity. If the
set of expectation values $\{ \tra(M_i\rho) \}$ is tomographically
complete, obviously, the bound is promised to give the exact
value, but in practice, a much smaller number of measurements is
sufficient to arrive at good bounds. Data of expectation values
can be composed, that is if two sets of expectation values are
combined, the resulting bound can only become better, to the
extent that two sets that only give rise to trivial bounds can
provide tight bounds. The approach presented here is perfectly
suitable for any finite-dimensional system, and also for
continuous-variable systems, as long as the observables $M_i$ are
bounded operators. Photon counting with a phase reference gives
rise to such operators, as we will see. Note also that similar
ideas, formulating lower bounds to entanglement measures,
constraining expectation values of observables can also be
formulated for other measures of entanglement \cite{eba07,grw07}.
This is in line with the idea of systems identification of trying to
directly estimate relevant quantities, instead of aiming at the 
detour of reconstructing the quantum states first.

\section{Photon-number-resolved weak homodyne detection}
\label{sec:PNR}

%%%%IAW-091106%%%%
We consider now the application of entanglement quantification to
detection of entangled photonic states. In this application we
propose to make use of photon-number resolving detectors. These
have several useful features that make them well-suited to the
measurement of non-classical signatures of light beams. First,
weak-field homodyne detection provides a way to demonstrate the
entangled character of EPR-like two-mode squeezed
states~\cite{Grangier88, Kuzmich00, Banaszek9899}, in contrast to
strong-field homodyne detection~\cite{Ou92}. Second, because the
amplitude of the local oscillator is comparable to that of the
signal, the phase sensitivity of the photon counting distribution
is much smaller than that of a regular homodyne detector.~This
greatly reduces the problem of synchronizing local reference
frames, which generally becomes more difficult with increasing
distance.

Within the framework of the quantum theory of measurement, the
action of a detector is completely  specified by its positive
operator valued measurement (POVM) set \cite{Holevo}. A POVM
element is a positive definite operator $ \Pi_{\beta, \gamma}\geq
0$, which represents the outcome $\beta$ of a given detector, for
a setting $\gamma$ corresponding to a particular value for a
tunable parameter in the detector. In the case of a homodyne
detector, $\gamma$ would correspond to the phase or the amplitude
of the local oscillator. The complete set should satisfy
$\sum_{\beta} \Pi_{\beta, \gamma}=\mathbb{I}$. The probability
$p_{\beta, \gamma}$ of obtaining outcome $\beta$ for setting
$\gamma$ can be related to the state of the system $ \rho$ by
$p_{\beta, \gamma}=\tra( \rho  \Pi_{\beta, \gamma})$.

Our detection scheme consists of a weak local oscillator (LO)
mixed with the signal $ \rho$ at a variable reflectivity ($R$)
beam-splitter (BS). The outcome of such an interference is
collected by time-multiplexed photon-number-resolving (PNR)
detectors~\cite{Achilles}. The time-multiplexed detectors (TMD)
split the incoming pulses into $8$ distinct modes, which are
eventually detected by binary avalanche photo-diodes (APDs) which
can register either $0$ or $1$ click. Thus there are $9$ possible
outcomes for a given TMD which are labelled by the number of
clicks $\beta= 0,\dots,8$. The settings of the detector $\gamma$,
in turn, are determined by a number of experimental parameters,
such as LO amplitude $|\alpha|$ and phase $\theta$, BS
reflectivity $R$, and detector efficiency $\eta$. By tuning the
detector settings it is possible to prepare POVM elements able to
project onto a large variety of radiation field states, ranging
from Fock states to quadrature squeezed states \cite{Puentes}.

%In order to obtain an expression for $ \Pi_{\beta, \gamma}$ it is
%necessary to
%relate the probability of a given outcome $p_{\beta, \gamma}$ with
%the input state $\rho$, for the specific detection system. This
%is described in the next subsection.

\subsection{Detector model}

\begin{figure}[h!]
\begin{center}
\includegraphics[angle=0,width=10truecm]{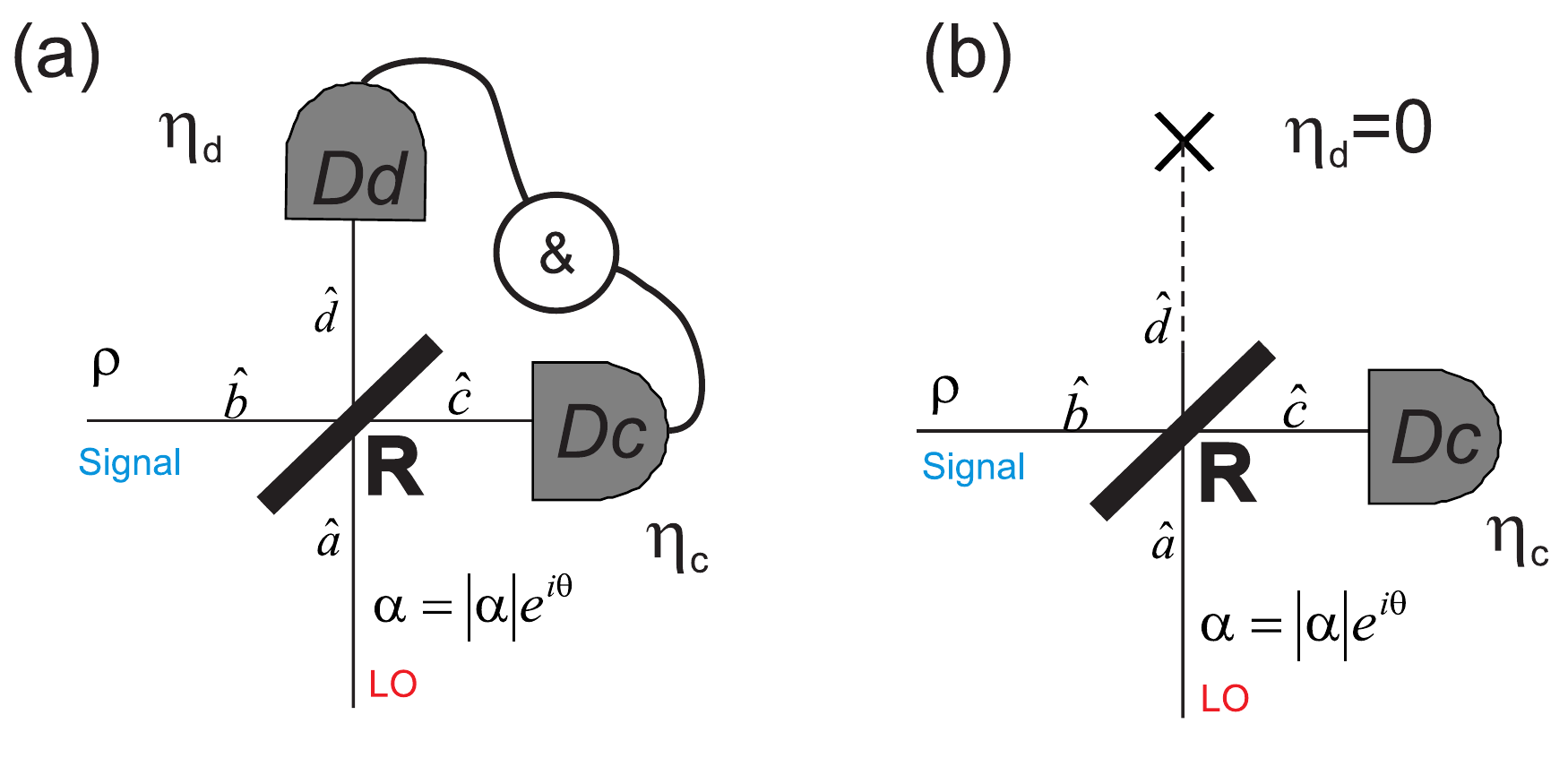}
\caption{Homodyne detection scheme for (a) balanced and (b)
unbalanced configuration. $D_{c,d}$ are PNR detectors of the
time-multiplexed type.} \label{fig:1}
\end{center}
\end{figure}

In Fig.~(\ref{fig:1} (a)) we show a schematic of our detection
system for the balanced configuration. The BS input modes, labeled
$a$ and $b$, correspond to the LO and the signal ($ \rho $),
respectively. The output modes, labeled by $c$ and $d$, are
detected by PNR detectors $D_{c}$ and $D_{d}$. The joint detection
events, denoted $\{\beta=(n_{c},n_{d})\}$, are recorded for
different LO settings $\gamma=(|\alpha|,\theta)$. The LO, is
prepared in a coherent state with state vector $|\alpha \rangle $ of complex
amplitude $\alpha=|\alpha|e^{i\theta}$ and provides the phase
reference needed to access off-diagonal elements in $ \rho$, as
the PNR detectors alone have no phase sensitivity \cite{Achilles}.
For ideal PNR detectors, the probability to obtain measurement
outcome $\beta$ for LO setting $\gamma $ is related to $\rho$
by \cite{Pregnell}

 \begin{equation}
 \label{eq:3}
 p_{\beta, \gamma }=\mathrm{Tr}_{c,d}[{U}\sigma{U}^{\dagger } |n_c, %%@
n_d\rangle \langle n_c, n_d|],
 \end{equation}%

\noindent where $U=e^{i\chi(b^{\dagger}a+a^{\dagger}b)}$ is the unitary operator %%@
representing the BS, $R=\cos(\chi)^2$ is the BS reflectivity, $ %%@
\sigma
=|\alpha \rangle \langle\alpha |_a \otimes  \rho_b$ the two-mode
input state and $|n_c, n_d \rangle=|n_c\rangle_c |n_d\rangle _d$
the photon number state vectors of mode $c$ ($d$) to be detected at PNR
detectors $D_{c}$ ($D_{d}$).

In order to account for the imperfections of the time-multiplexed
PNR detectors we use a well tested model of the TMDs
\cite{Achilles}. Within this model, the TMD operation can be
described as  a map from the incoming photon-number distribution
$\vec{r}$, as a vector (i.e., the diagonal components of the
density matrix) to the measured click statistics $\vec{k}$ by
$\vec{k} =C \cdot L\cdot \vec{r} $.~Here $L$ and $C$ are matrices
accounting for loss and the intrinsic detector structure
\cite{Achilles}, respectively. To calculate the POVM elements
implemented by our  PNR homodyne detector, the POVMs for TMD
detectors $D_c$ and $D_d$ are determined from the $C$ and $L$
matrices (characterized by independent methods \cite{detectomo}).
The TMD POVMs are then substituted into Eq.\ (\ref{eq:3}), in
place of the photon number projectors $|n_c, n_d\rangle \langle
n_c, n_d|$, to obtain the final expression for the imperfect POVM
elements $ \Pi _{\beta,\gamma }$. We note that our TMDs can
resolve up to $8$ photons, setting the number of possible outcomes
to $81$.

\subsection{Unbalanced detection scheme}

Our aim is to use such homodyne PNR detectors to provide lower
bounds on the entanglement of bipartite quantum states, in which
case two of such devices should be employed. To this end, the
joint POVM statistics of the four modes involved in the detection
need to be measured, increasing the total number of  POVM elements
to $81^2=6561.$ In order to simplify the experimental arrangement,
we use the detector in an unbalanced configuration, so that we
only detect one of the outgoing modes of each homodyne BS. In this
way only two modes need to be jointly detected and the total
number of POVM elements is reduced to $9^2=81$. This unbalanced
scheme can be modeled by setting the efficiency $\eta$ of one the
PNR detectors to zero (see Fig.~(\ref{fig:1} (b))). The only
disadvantage of this unbalanced scheme is that the overall
efficiency is in principle reduced by $50\%$,  but this limitation
can be overcome by increasing the BS reflectivity $R$.
Additionally, as we will show in the next section, our partial
detection approach alleviates the strong efficiency requirements
of full tomography allowing for the additional losses of the
unbalanced scheme.

\begin{figure}[h!]
\begin{center}
\includegraphics[angle=0,width=9truecm]{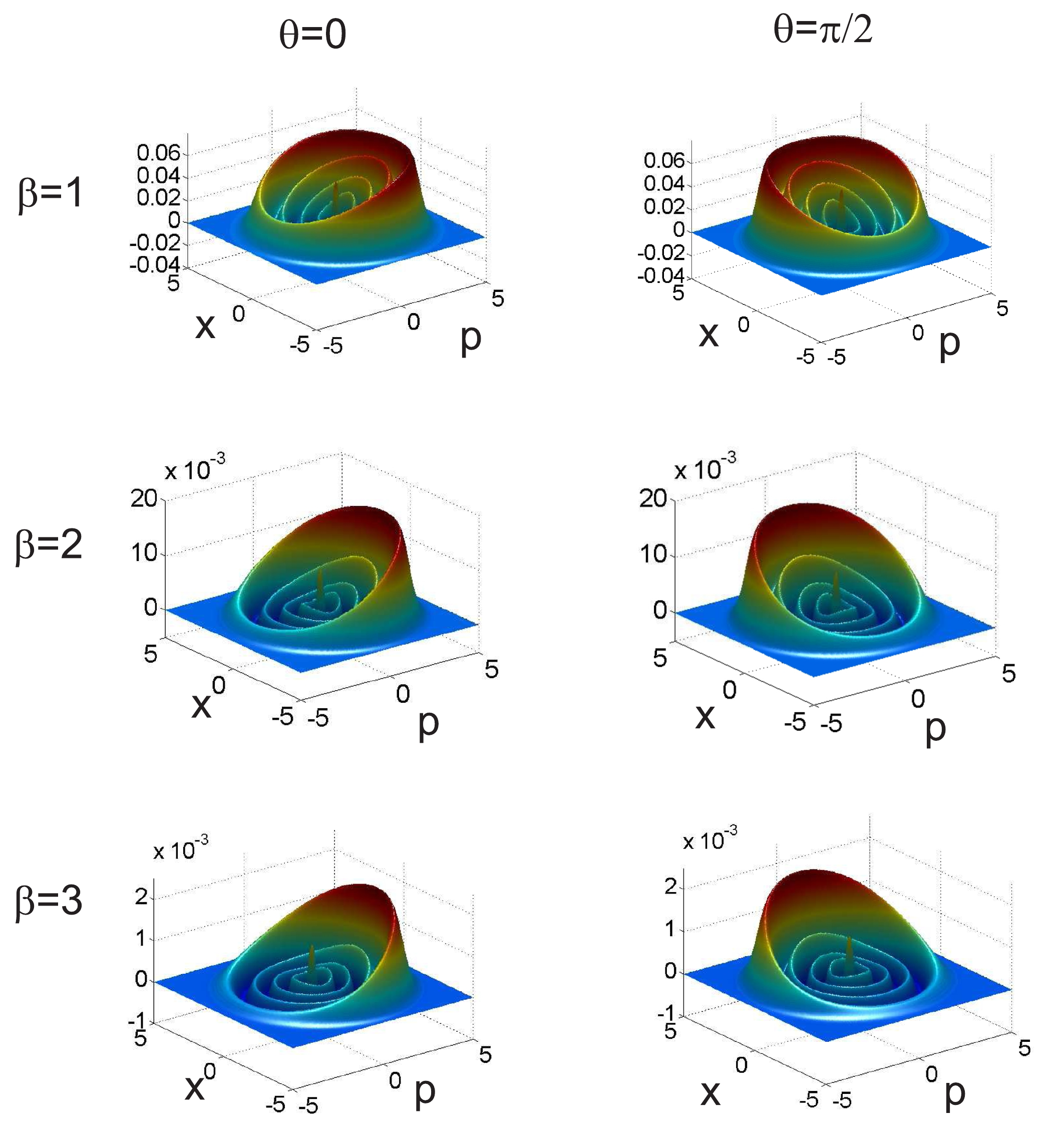}
\caption{Wigner representation of selected POVM elements $
\Pi_{\beta, \gamma}$ for the unbalanced scheme described in
Fig.~(\ref{fig:1} (b)). Different columns correspond to LO phase
settings $\theta=0$ and $\theta=\pi/2$. The rows correspond to
three different detector outcomes $\beta=1,2,3$. The amplitude
of the LO, BS reflectivity and detector efficiency are,
$|\alpha|=1$, $R=50\%$ and $\eta=10\%$ , respectively.}
\label{fig:2}
\end{center}
\end{figure}

In Fig.~(\ref{fig:2}), we show numerically constructed Wigner
functions corresponding to $6$ different POVM elements $ \Pi_{\beta,
\gamma}$, characterizing the unbalanced scheme. The axes $(x,p)$
label the phase space quadratures. The different columns
correspond to different LO phases $\theta =0$ and $\theta= \pi/2$.
The rows correspond to three consecutive outcomes
$\beta=1,2,3$, labelling the corresponding number of detector
clicks. For these simulations we fixed the amplitude of the LO to
$|\alpha|=1$, the BS reflectivity to $R=50\%$ and the detector
efficiency to $\eta=10\%$, which is a realistic value for a single
mode TMD. The figure shows that $ \Pi_{\beta, \gamma}$ are not
rotationally symmetric, as expected for a phase sensitive
detector. The oscillations in the Wigner functions are due to the
low efficiency of the detectors which mixes  different photon
number states, whose phase-space representation is given by
consecutive rings of increasing radii.  Also, as is expected, a
change in the LO phase setting  by $\pi/2$ corresponds to an
overall phase-space rotation in the Winger function. In the next
section we show that by using $8$ POVM elements of the type shown
in Fig.~(\ref{fig:2}) for each subsystem, we can construct the
measurements $M_i$ mentioned in Sec.~(\ref{sec:SDPs}), which can
be employed to bound the entanglement content of two-mode
degaussified states.

\section{Application to continuous-variable entanglement %%@
distillation}
\label{sec:CVentdist}

We will now apply the above methods to a setting that plays a
central role in continuous-variable entanglement distillation.
Entanglement distillation aims at producing more highly entangled
states out of a situation where entanglement is present only in a
dilute and noisy form, presumably generated by some lossy quantum
channel. It provides the key step in quantum repeater ideas
allowing for the distribution of long-range entanglement in the
presence of noise. Crudely speaking, one may distinguish between
actual distillation schemes that involve more than one specimen of
an entangled state at each step of the protocol, and
``Procrustean'' or local filtering approaches that take a single
copy of a state and under appropriate local filtering give rise --
if successful -- to a more highly entangled state. In the setting
of strict Gaussian operations, continuous-variable entanglement
distillation of neither kind is possible \cite{distillation}, but
this obstacle can be overcome with the help of non-Gaussian
ingredients such as photon addition or subtraction \cite{Eisert}.
Such first Procrustean steps can also be used as starting points
in full entanglement distillation protocols. Indeed, quite
exciting first steps towards full continuous-variable entanglement
distillation have recently been taken experimentally
\cite{Ourjoumtsev07,Dong,Takahashi,Xiang,Bellini}.

In the subsequent discussion we show the use of quantitative tests to certify
success in such a scheme. Needless
to say, we discuss specific input states, but it should be clear that the
given entanglement bounds do not make use of that a-priori knowledge.
We consider as our initial state vector $|\psi^{\mathrm{ini}}\rangle$ an
ideal pure two-mode quadrature squeezed state of the form
 \begin{equation}
 \label{eq:9}
 |\psi^{\mathrm{ini}}\rangle =  %%@
\sqrt{1-\lambda^2}\sum_{n=0}^\infty\lambda^{n}|n,n\rangle_{1,2},
 \end{equation}
\noindent where $\lambda$ represents the squeezing parameter, and the
subindices $(1,2)$ represent each spatial mode. Such type of
states are produced in the laboratory by the non-linear process of
spontaneous parametric down conversion (SPDC) in non-linear
crystals \cite{Mosley}. In order to simplify our numerical
calculations, we will restrict the maximum photon-number per mode to $n_{\mathrm{max}}=3$.
Thus the set of bipartite initial states $\rho_{1,2}^{\mathrm{ini}}$  %%@
is given by the set of $16\times16$
density matrices $\rho_{1,2}^{\mathrm{ini}}=
|\psi^{\mathrm{ini}}\rangle\langle\psi^{\mathrm{ini}}|$.
The Logarithmic Negativity for the bipartite state in
Eq.\ (\ref{eq:9}) takes the simple form
 \begin{equation}
\mathcal{N}(\rho_{1,2}^{\mathrm{ini}})=\log\|(\rho_{1,2}^{\mathrm{ini}}%%@
)^{T_1}\|_1=\log\left(\frac{1+\lambda}{1-\lambda}\right),
 \end{equation}
 as can readily be verified.

\begin{figure}[h!]
\begin{center}
\includegraphics[angle=0,width=12truecm]{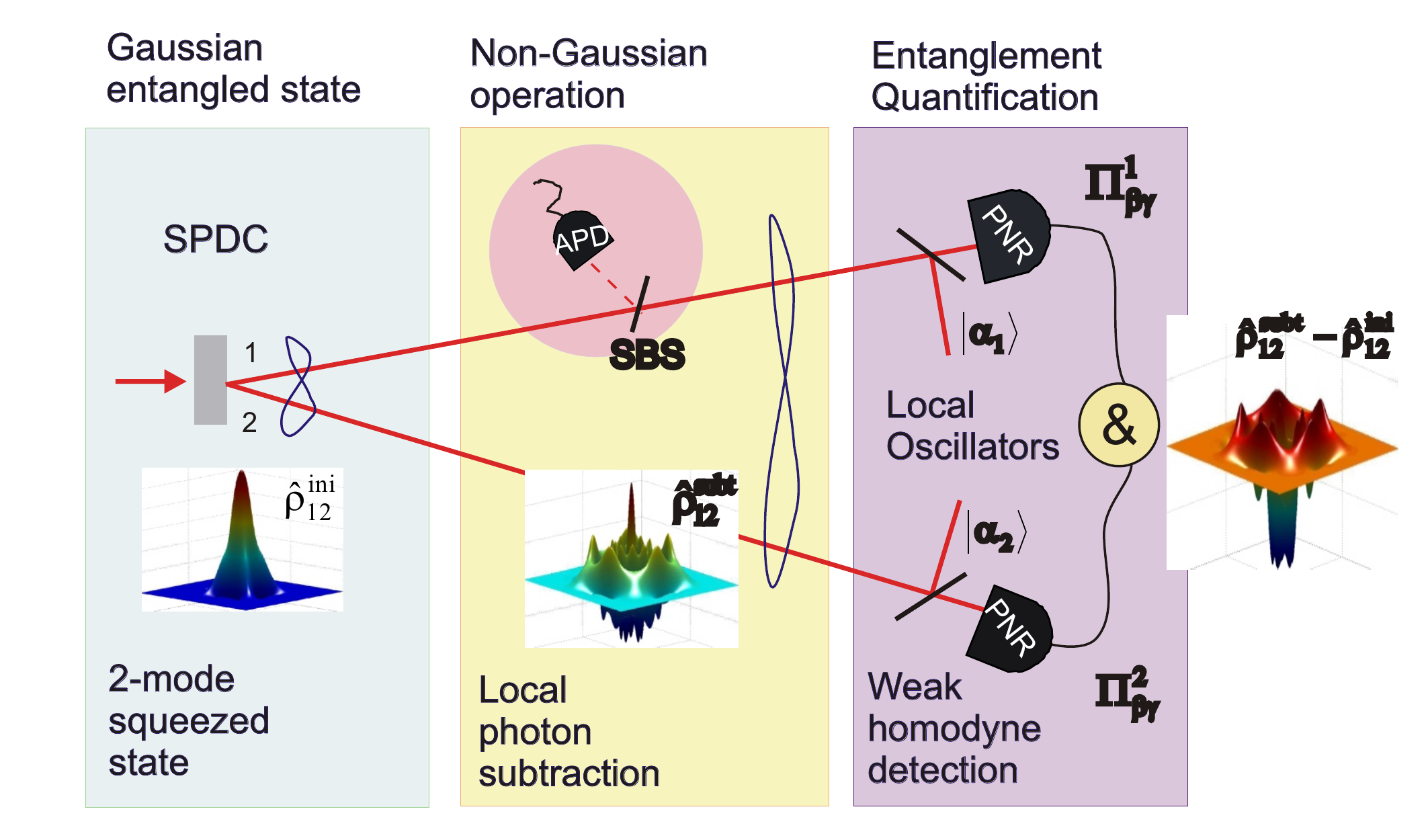}
\caption{(Color online) Scheme describing the bipartite initial
state $ \rho_{1,2}^{\mathrm{ini}}$ produced by spontaneous
parameteric down conversion (SPDC). Next, the poton-subtracted
state $ \rho_{1,2} ^{\mathrm{subt}}$ is prepared by local
subtraction of a single photon at a tunable subtraction beam splitter %%@
(SBS).
The entanglement content in  $\rho_{1,2} ^{\mathrm{subt}}$ is %%@
quantified by our partial detection approach. This type of  scheme is %%@
one of the main components of an
entanglement distillation protocol \cite{Eisert}. \label{fig:7}}
\end{center}
\end{figure}

\subsection{Two-mode photon-subtracted quadrature squeezed state}

In order to distill continuous-variable entanglement from Gaussian %%@
states, such as the two-mode quadrature squeezed state  described by %%@
Eq.\ (\ref{eq:9}), an operation that removes the Gaussian nature of %%@
the probability distribution is required \cite{Eisert}.~Examples of %%@
such non-Gaussian operations are the conditional subtraction or  %%@
addition of a  photon \cite{Ourjoumtsev07,Bellini}.~An ideal %%@
two-mode photon-subtracted quadrature squeezed state can be
modeled by inserting a BS of transmission $T$ (the so-called
subtraction beam-splitter SBS) in one spatial mode. The reflected
mode from the SBS is then detected by a standard (ideal) avalanche
photodetector (APD) (this is schematized in Fig.~(\ref{fig:7})).
The photon subtracted state can, in the approximation of having a
a very weakly reflecting beam-splitter, thus be written as
 \begin{equation}
\label{eq:substate}
 |\psi^{\mathrm{subt}}\rangle=C\sum_{n=1}^{n_{\max}}(\lambda %%@
T)^{n}\sqrt{n}|n-1,n\rangle_{1,2},
 \end{equation}
where $C$ is a normalization constant and in our simulations
$n_{\mathrm{max}}=3$. The corresponding density matrix is
$\rho_{1,2}^{\mathrm{subt}}=|\psi^{\mathrm{subt}}\rangle\langle\psi
^{\mathrm{subt}}|$. We note that the family of states described by
Eq.\ (\ref{eq:substate}) are of current interest in the realm of
continuous-variable entanglement distillation, as they represent a
particular kind of non-Gaussian state (i.e., a state whose Wigner
representation is not Gaussian), whose entanglement content
$\mathcal{N}(\rho^{\mathrm{subt}}_{1,2})$ is predicted to be larger
than $\mathcal{N}(\rho^{\mathrm{ini}}_{1,2})$, for suitable
experimental parameters ($\lambda$, $T$)~\cite{distillation}.

\subsection{Construction of the observables}

Our aim is to construct bipartite measurement operators $M_i$ as a
tensor product of the POVM elements corresponding to each
subsystem $ \Pi_{1}\otimes  \Pi_{2}$. In particular, we will
consider $8$ POVM elements for each subsystem $(1,2)$, specified
by four different outcomes $\beta =0,1,2,3$ and two different
settings $\gamma$ corresponding to $\theta=0$ and $\theta=\pi/2$.
Thus the selected POVM subset for each mode consist of $8$ elements
$ \Pi_{\beta, \gamma}$, collected as
 \begin{equation}
  \{ \Pi_{0,0},\Pi_{1,0},\Pi_{2,0}, \Pi_{3,0}, %%@
\Pi_{0,\pi/2}, \Pi_{1,\pi/2},\Pi_{2,\pi/2}, \Pi_{3,\pi/2}\},
 \end{equation}
where we will keep this ordering in the POVM elements for the rest
of the paper. We measure $8\times8$ configurations, which
determine $64$ POVMs $M_{i}$, labelled by the index $i$, of the form
in Eq.\ (\ref{eq:localmeas}) with $j,k=1,\dots,8$ being the indices
labelling the POVM elements of mode $(1,2)$ respectively, and
$i=(j,k)$ the joint index, labelling the bipartite measurement
operator. For example, the observable $ M_{(6,1)}$ corresponds to
the POVM elements $\Pi_{1,\pi/2}$ for mode 1 and $\Pi_{0,0}$  for
mode $2$. This gives a total of 64 measurements, which in turn determine 64 expectation
values $\tra( \rho_{1,2} M_{i})=m_{i}$, with $i=1,\dots,64$. This is
a clear reduction with respect to full state tomography, which
would require (at least) $16^2-1=255$ measurements in order to
reconstruct $ \rho_{1,2}$ in a truncated Hilbert space of dimension
$16$. 

In order to find the lower bound on the Logarithmic Negativity of
the photon-subtracted states $ \rho_{1,2}^{\mathrm{subt}}$
described in Eq.\ (\ref{eq:substate}), by means of the set of
measurement observables $M_i$ as defined in
Eq.\ (\ref{eq:localmeas}), we follow the procedure described in
Sec.~(\ref{sec:SDPs}). Note that in a real experiment  $\tra(
\rho_{1,2}^{\mathrm{subt}} M_i)$ should
be replaced by the actual experimental probability estimates which will be subject
to different sources of noise. We will discuss the effect of
experimental noise on entanglement bounds in the final subsection.

\subsection{Numerical results}

\begin{figure}[h!]
\begin{center}
\includegraphics[angle=0,width=10truecm]{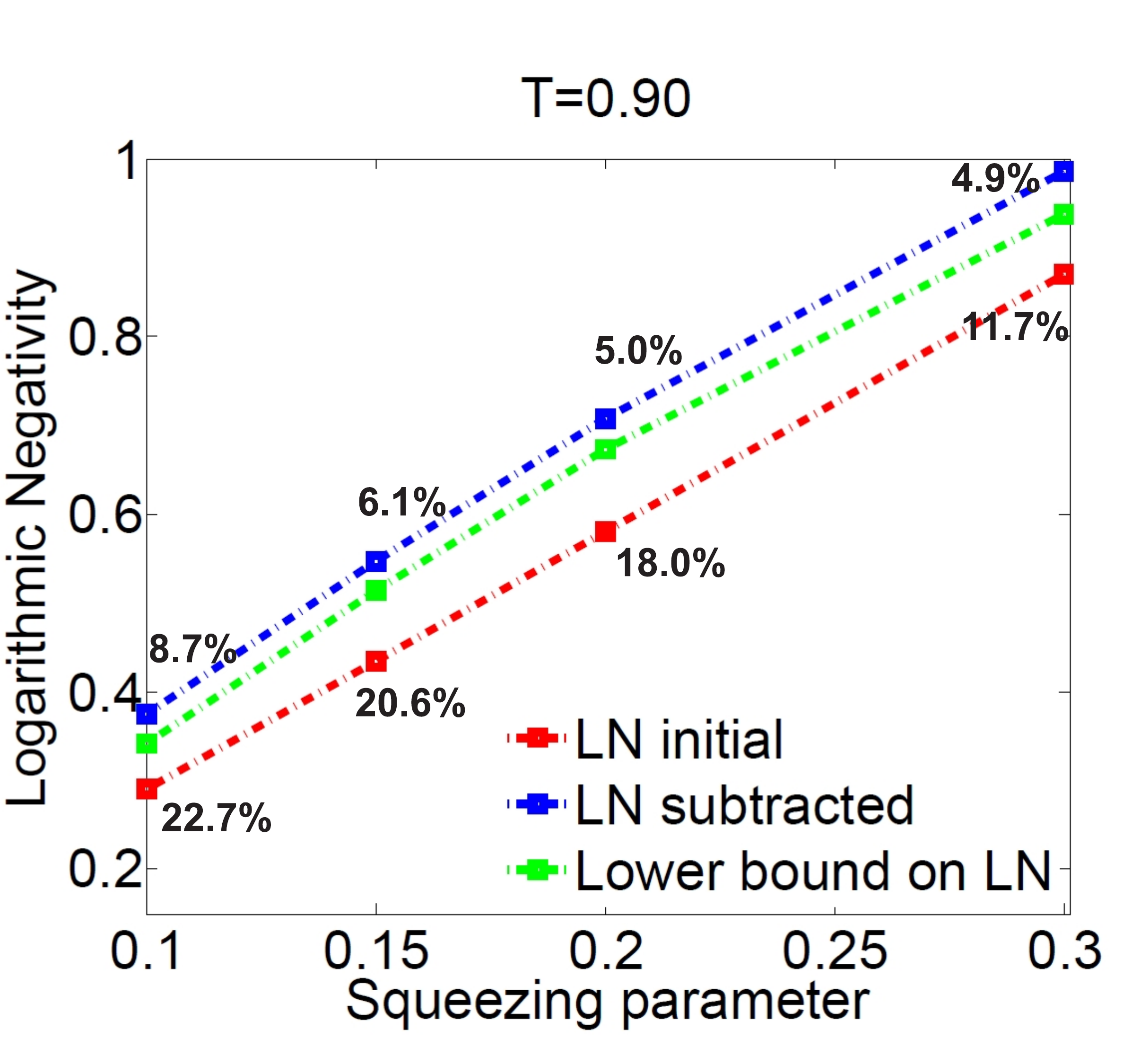}
\caption{(Color online) Logarithmic Negativity (LN) for the initial
state (red curve), subtracted state (blue curve) and lower bound on %%@
LN of subtracted state obtained by convex optimization (green curve), %%@
vs.\ squeezing
parameter ($\lambda$). Percentual entanglement increase with a
single photon-subtraction step and percentual difference between the %%@
actual LN of the subtracted state and the one obtained by convex %%@
optimization  are indicated. The SBS
transmission was fixed at $T=90\%$. \label{fig:3}}
\end{center}
\end{figure}

\begin{figure}[h!]
\begin{center}
\includegraphics[angle=0,width=10truecm]{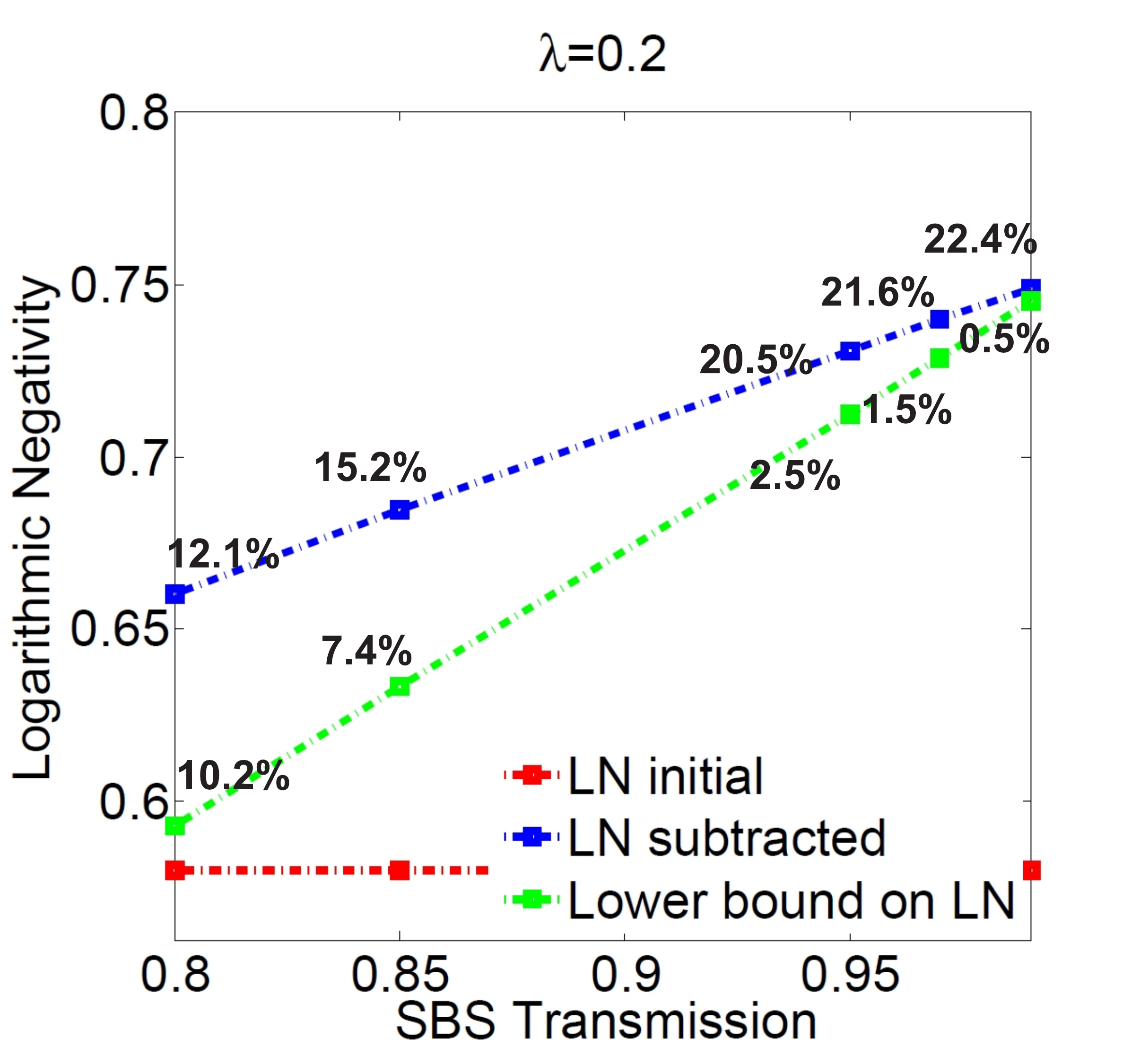}
\caption{(Color online) Logarithmic Negativity (LN) for the initial
state (red curve), subtracted state (blue curve) and lower bound on %%@
LN of subtracted state obtained by convex optimization (green curve), %%@
vs.\ SBS transmission ($T$). Percentual entanglement increase with a
single photon-subtraction step and percentual difference between the %%@
actual LN of the subtracted state and the one obtained by convex %%@
optimization  are indicated. The squeezing parameter was fixed at %%@
$\lambda=0.2$. \label{fig:4}}
\end{center}
\end{figure}

Fig.~(\ref{fig:3})  presents the percentual entanglement increase
between $|\psi^{\mathrm{ini}}\rangle$ (red curve) and
$|\psi^{\mathrm{subt}}\rangle$ (blue curve) for different
squeezing parameters $\lambda$ ranging from $\lambda=0.1$ to
$\lambda=0.3$. Percentual differences between the actual
Logarithmic Negativity characterizing the photon-subtracted state
and the lower bound obtained by convex optimization (green curve)
are also indicated. The transmission coefficient of the
subtraction beam-splitter (SBS) in Fig.~(\ref{fig:2}) was fixed at
($T=90\%$), the LO amplitude and detector efficiency were set to
$|\alpha|=1$ and $\eta=10\%$, respectively.  It is noticeable that
while  a single-photon-subtraction step produces a larger
entanglement increase for lower values of $\lambda$, the lower
bound on the entanglement becomes tighter for higher squeezing
parameter $\lambda$. The percentual error in the lower bound is in
all cases below $9\%$, which reveals the accuracy of our partial
detection scheme in characterizing entanglement.

Fig.~(\ref{fig:4}) presents the percentual entanglement increase
between $|\psi^{\mathrm{ini}}\rangle$ (red curve) and
$|\psi^{\mathrm{subt}}\rangle$ (blue curve) for different SBS
transmission $T$, ranging from $T=80\%$ to $T=99\%$. Percentual
differences between the actual Logarithmic Negativity
characterizing the photon-subtracted state vector
($|\psi^{\mathrm{subt}}\rangle$) and the lower bound obtained by
convex optimization (green curve) are also indicated. The
squeezing  parameter was fixed at ($\lambda=0.2$), the LO
amplitude and detector efficiency were set to $|\alpha|=1$ and
$\eta=10\%$, respectively. Fig.~(\ref{fig:4}) shows that for a
fixed squeezing parameter the single-photon-subtraction step
produces a larger entanglement increase for higher SBS
transmission $T$ and that the lower bound becomes tighter for
higher $T$.~In all cases, the percentual error in the lower bound
remains below $11\%$.

\begin{figure}[h!]
\begin{center}
\includegraphics[angle=0,width=13.5truecm]{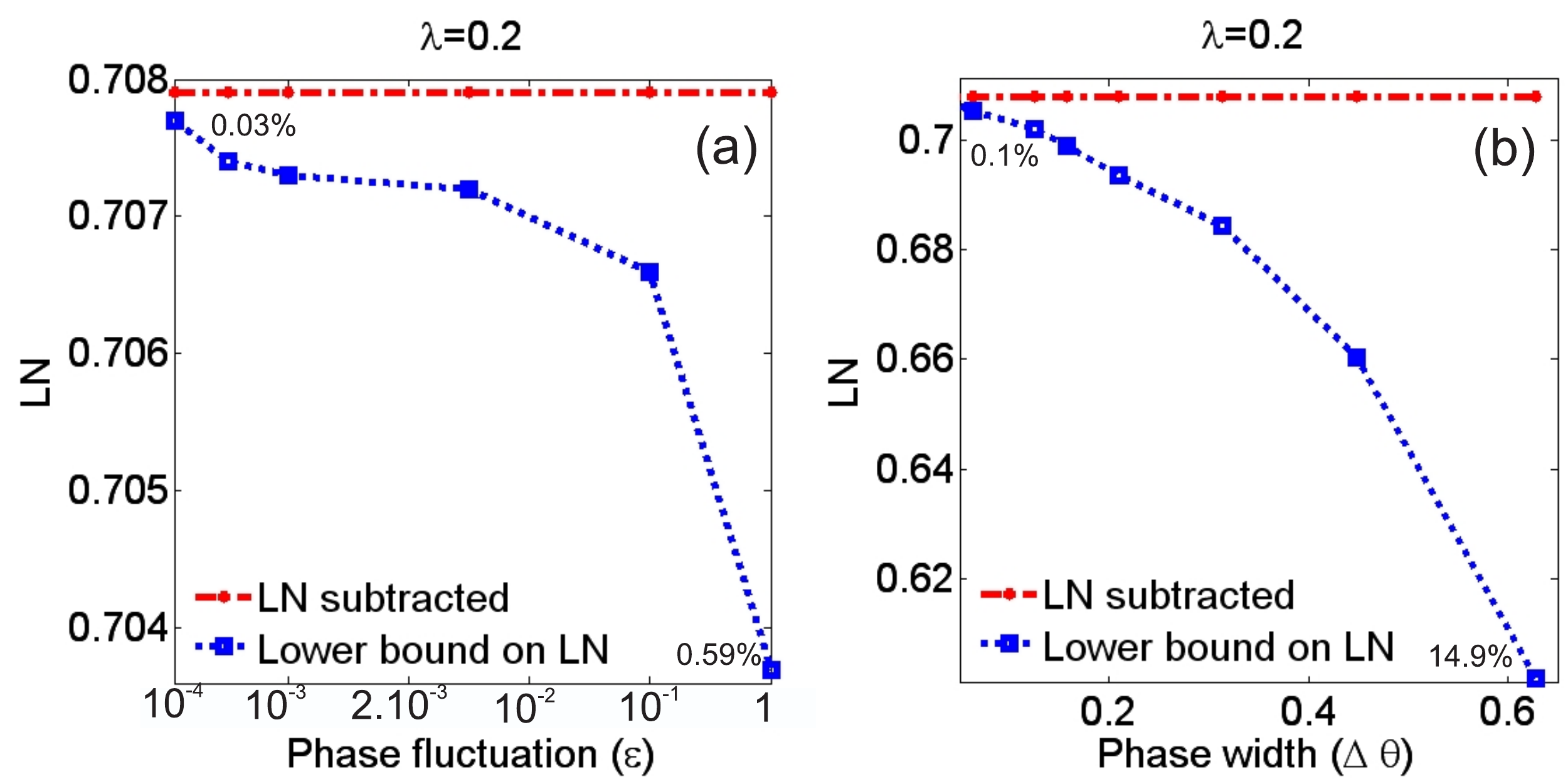}
\caption{(Color online) Exact Logarithmic Negativity (red curve)
and  lower bound obtained by convex optimization (blue curve) for
the photon-subtracted state  vs.\ (a) dimensionless phase noise
$\epsilon$ and (b) phase noise standard deviation $\Delta \theta$ in radians for LO
phase settings $\theta_0=(0,\pi/2)$. Max. and min. percentual
differences are indicated. The squeezing parameter was fixed at
$\lambda=0.2$.\label{fig:5} }
\end{center}
\end{figure}

Next, we tested the robustness of the measurement scheme with
respect to different types of LO phase noise. To this end we
constructed a set of 64 POVM elements as described in the previous
section, for a fixed LO amplitude $\alpha=1$, detector efficiency
$\eta=0.10$, SBS transmission $T=0.90$ and squeezing parameter
$\lambda=0.2$. The two fixed phase setting $\theta_0=(0,\pi/2)$
were subject to different types of fluctuations. In particular, we
investigated the required precision in the LO phase  $\theta$ by
adding different amounts of random phase noise $\epsilon$ in the
form $\theta=(0\pm \epsilon/10, \pi/2(1 \pm \epsilon/10))$, where
$0 \leq \epsilon \leq 1$ is a random number with a uniform
distribution. In our numerical simulations we found that for a
phase error  of up to $10\%,$ the lower bound differs from the
actual Logarithmic Negativity by less than $1\%$. This means that
an LO phase precision of 5 degrees (at the most) is required for
the bounds to produce a highly tight estimate. This is shown in
Fig.~(\ref{fig:5} (a)), for $\theta_0=(0,\pi/2)$, a squeezing
parameter $\lambda=0.2$, a SBS transmission $T=90\%$, an LO
amplitude $|\alpha|=1$ and a detector efficiency $\eta=0.10$.

We also analyzed the effect of temporal phase fluctuations, by
modelling the LO as a phase averaged coherent state described by
the complex amplitude $\alpha=\sum_{j} |\alpha| e^{i(\theta_0 +
\delta \theta_j)}$ with $j=1,\dots,100$ and $\delta \theta$ a random
phase with a uniform distribution centered around
$\theta_0\in[0,\pi/2]$ and with width $\Delta \theta$. We found that
a phase width of up to $0.6$ radians ($\approx 30$ degrees)
introduces a percentual difference in the lower bound of up to
$15\%$. For a phase width $\Delta \theta$ of up to $0.4$ radians
($\approx 20$ degrees) the lower bound on the logarithmic
negativity is within $10\%$. This is shown in Fig.~(\ref{fig:5}
(b)), for a squeezing parameter $\lambda=0.2$, a SBS transmission
$T=90\%$, an LO amplitude $|\alpha|=1$ and a detector efficiency
$\eta=0.10$.

\begin{figure}[h!]
\begin{center}
\includegraphics[angle=0,width=14truecm]{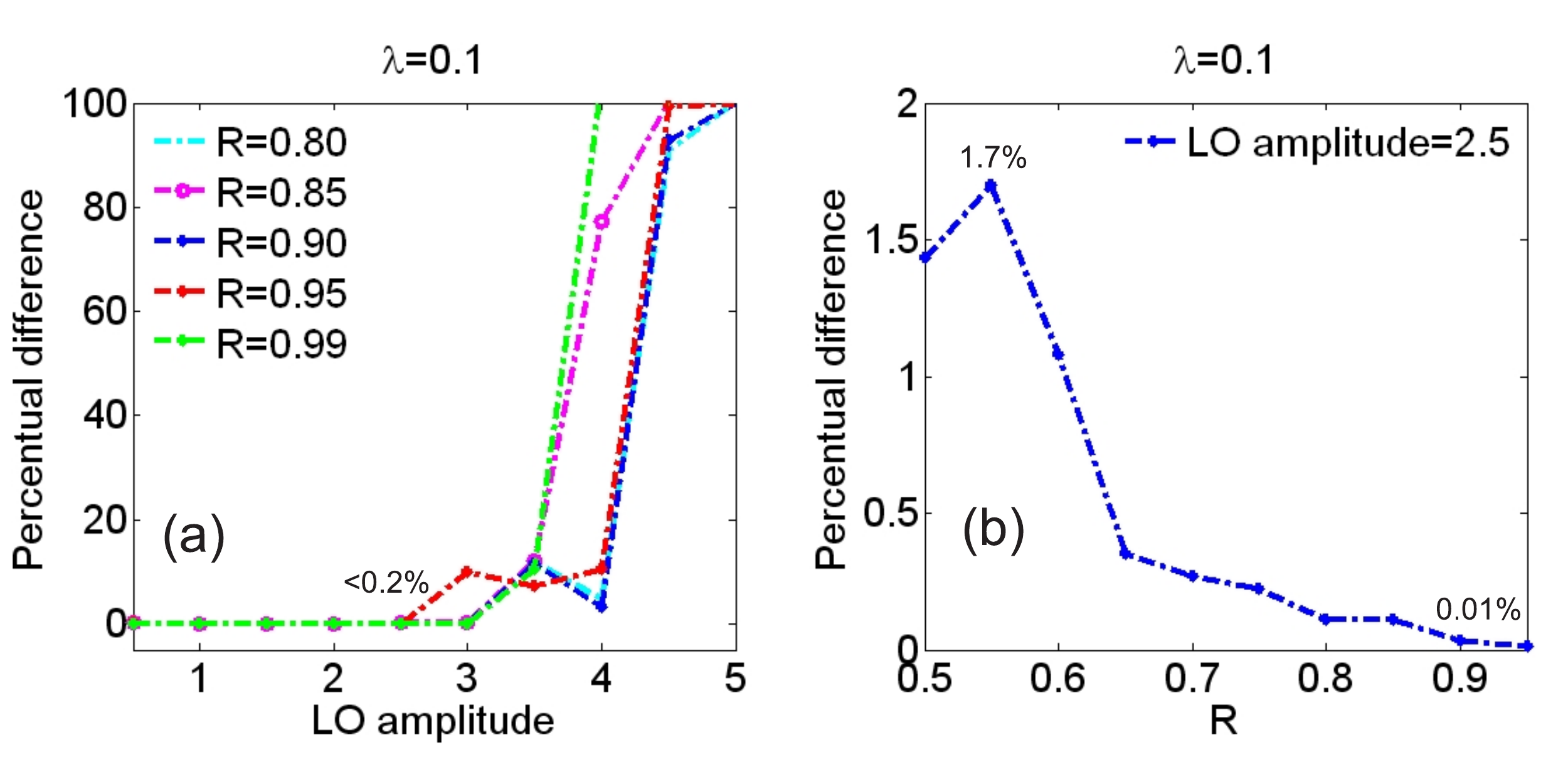}
\caption{(Color online) (a) Percentual error in the lower bound
set by convex optimization for different homodyne BS reflectivities %%@
$R$.
The error remains below $ 0.2 \%$ for a sufficiently weak LO
amplitude $|\alpha| < 2.5$. (b) Extension to a larger range of BS
reflectivities $R$, for $\lambda=0.1$, $|\alpha|=2.5$,
$\theta=(0,\pi/2)$.  \label{fig:6}}
\end{center}
\end{figure}

Finally, we analyzed the impact of a different homodyne BS
reflectivity $R$ on the overall accuracy of the entanglement %%@
quantification scheme. We found that for $R\geq 80\%$ the lower
bound on the Logarithmic Negativity differs by less than $0.2\%$
from the actual value, as long as the LO
amplitude remains small enough ($|\alpha|\leq2.5$) due to the
limited photon-number resolution in the time-multiplexed
detectors. This is shown in Fig.~(\ref{fig:6} (a)).
Fig.~(\ref{fig:6} (b)) shows a complete simulation for $50 \%\leq
R \leq 99 \%$, $|\alpha|=2.5$, $\lambda=0.1$ and $T=90\%$. In all
the simulations, the  TMD efficiencies were set to $\eta=0.10$ and
the LO phase settings were chosen as $\theta=(0,\pi/2)$.
Additionally, the subtraction APD in Fig.~(\ref{fig:7}) is assumed
to have a limited efficiency, which is modelled by interposing a
beamsplitter with transmittivity of $15\%$. The numerical
simulations in this work were  implemented using the convex
optimization package SeDuMi \cite{sedumi}.

\subsection{Tolerance to experimental measurement errors}
In the numerical simulations presented here we have used the exact
expectation values $m_i=\tra( \rho^{\mathrm{subt}} M_i)$ for the
minimization of Logarithmic Negativity. However, in a real
experiment such expectation values are affected by different
sources of noise. In this subsection, we test the  tolerance of
the scheme to experimental errors. There are several ways to
include such an error. One approach would be to estimate variances
of measured values, and then make a model including Gaussian-distributed
errors for the measured variables. Another would be a hard bound on the degree of
entanglement as a function of a small norm deviation from the
perfect data, giving rise to a box error model. This latter
error model even allows for a malicious correlation in the errors,
in that all errors add constructively.  Clearly, independent, identically distributed errors would give
rise to much more robust bounds.

Nonetheless, in order to  evaluate the feasible robustness of our method, we will now refer
to this latter, more demanding error model: We merely
require for an $\epsilon>0$
that the measured value $n_i$ and the true
expectation value $m_i= \tra{(M_i \rho)}$ satisfy $m_i \in [(1-\epsilon)n_i , (1+\epsilon)n_i]$
for all $i$.  Hence, the problem to be solved becomes
 \be
 \label{eq:Emin}
 \mathcal{N}_{\min}  = \min_{\rho}\{\mathcal{N}(\rho): \tra(\rho
 M_i)=n_i,\, m_i \in [(1-\epsilon)n_i , (1+\epsilon)n_i] \,\,\, \forall i \},
 \ee
Including such measurement errors,
Eq.\ (\ref{eq:sdp}) then clearly becomes
 \ben
 \label{eq:exptsdp}
 &&\mathrm{maximize}\;\;\;\; \log\Big(\sum_i \nu_i n_i \Big), \\
 &&\mathrm{subject\;to}\;\;H^{T_1}\geq \sum_i \nu_i M_i ,\nonumber\\
 &&  \hspace{2.0cm} m_i \in [(1-\epsilon)n_i , (1+\epsilon)n_i] \,\, \forall i,\nonumber\\
 && \;\;\;\;\;\; \mathrm{and}\;\;\;   -\mathbb{I} \leq H \leq
\mathbb{I},\nonumber
 \een
which can be solved as easily as Eq.\ (\ref{eq:sdp}) using
SeDuMi~\cite{sedumi}. Note that the resulting bound is even valid
if each of the errors in the measured data are maliciously correlated.

In this subsection, we use as an example a
two-mode squeezed state with $\lambda=0.2$ from which a photon is
subtracted using a SBS with $T=95\%.$ The APD in
Fig.~(\ref{fig:2}) is assumed to have a limited efficiency, which
is achieved by interposing a beamsplitter with transmittivity of
$20\%.$ In a short table below, we present the bounds attained by
solving the SDP in Eq.\ (\ref{eq:exptsdp}) for some representative
values of $\epsilon.$
\begin{center}
\begin{tabular}{|c|c|c|c|c|}
  \hline
  $\epsilon$ & 0.0 & 0.001 & 0.01 & 0.1 \\
  \hline
  $\mathcal{N}_{\min}$ & 0.7308 & 0.7185 & 0.6660 & 0.3034\\
  \hline
\end{tabular}
\end{center}
These numbers must be compared with the entanglement of the
initial two-mode squeezed state, which has $\mathcal{N}=0.5803,$
and the ideal photon subtracted state which has
$\mathcal{N}=0.7309.$ Note that the state on which we put the
lower bounds is inevitably mixed, and the table shows the
robustness and effectiveness of our scheme. $\epsilon =0.01$ is
enough to demonstrate the enhancement of entanglement by
distillation with experimentally realistic parameters, without
having to undertake a full tomography of the quantum states
involved. This value of $\epsilon$ translates to about $10000$
data points, via the central limit theorem, for each measurement
configuration. This is in line with the number of data points
taken in other experiments involving reconstruction of
non-Gaussian states~\cite{Ourjoumtsev07}.

\section{Conclusions}
\label{sec:conc}

We have presented quantitative numerical evidence that a novel homodyne detection scheme with photon-number resolution is able to set accurate bounds on the entanglement content of a
family of two-mode photon-subtracted quadrature squeezed states.
The entanglement lower bounds retrieved by the measurement scheme are
accurate to within $10\%$ for the full range of squeezing
parameters $\lambda=0.1-0.3$ and  subtraction beam-splitter
transmissions $T=80\%-99\%$. We found that the bounds become
tighter for higher $\lambda$ and $T$. We also analyzed the
required phase precision and stability in the local oscillator
(LO), and found that a precision of less than $5 $ degrees is
required for a bound accuracy within $1\%$, while  temporal phase
fluctuations of up to $20$ degrees can be accepted for a lower bound with $10\%$ accuracy. Additionally we found
that a homodyne beam-splitter reflectivity $R$ above $60\%$, for
an LO amplitude within $|\alpha|=2.5$ is sufficient to obtain a
lower bound on the Logarithmic Negativity which agrees to within
$2\%$ with the actual Logarithmic Negativity value characterizing
the photon-subtracted state. The results reported here provide
strong numerical evidence of the suitability of our partial detection
scheme for entanglement quantification of bipartite degaussified
states. We note that this type of partial detection approach is
not only attractive due to its accuracy but also due to its
scalability. This is of importance for the application of an
entanglement distillation protocol combining two degaussified
sources \cite{Eisert, distillation}.~In particular, our
scheme can be easily scaled to the detection of four spatial modes, in which case it would require the
measurement of only $64^2=4096$ outcome probabilities. In contrast,  full state
tomography would require (at least) $16^4 -1=65535$ different measurements. Therefore our method
provides a feasible, direct and resilient way of accurately
experimentally characterizing entanglement in continuous-variable
quantum systems. Finally, we anticipate the amount of data required in order to obtain
an adequate precision in the measurement-outcome probabilities characterizing our partial measurement scheme to be considerably lower than that required for full state tomography.

\section*{Acknowledgments}

This work was supported by the EPSRC through the QIP IRC, the EU through the IST directorate FET Integrated Project QAP, and through STREP projects CORNER, HIP, COMPAS and MINOS. JE acknowledges an EURYI Award, MP and IAW  Royal Society Research Merit Awards, and MP an
Alexander von Humboldt Professorship.

\end{document}